\documentstyle[aps,amsmath,amsfonts,amssymb,multicol,graphicx]{revtex}
\newcommand{\wb}{\omega_{\mathrm{b}}}

\begin{document}

\title{Breathers in a system with helicity and dipole interaction.}

%\author{B. S\'anchez-Rey}
%\thanks{Corresponding author. Email: bernardo@us.es}
%\address{Departamento de F\'{\i}sica Aplicada I, Universidad de
%Sevilla. Virgen de \'Africa 7, 41011-Sevilla, Spain}
\author{B. S\'anchez-Rey, JFR Archilla, F Palmero} \address{Departamento
de F\'{\i}sica Aplicada I, Universidad de  Sevilla.
 Av. Reina Mercedes s/n, 41012-Sevilla, Spain}
\author{FR Romero} \address{Facultad de F\'{\i}sica, Universidad de
Sevilla. Avda. Reina Mercedes s/n, 41012-Sevilla, Spain} %author

\date{Mars 2002}

\maketitle
\begin{abstract}
Recent papers that have studied variants of the Peyrard-Bishop
model for DNA, have taken into account the long range interaction
due to the dipole moments of the hydrogen bonds between base
pairs. In these models the helicity of the double strand is not
considered. In this particular paper we have performed an analysis
of the influence of the helicity on the properties of static and
moving breathers in a Klein--Gordon chain with dipole-dipole
interaction. It has been found that the helicity enlarges the
range of existence and stability of static breathers, although
this effect is small for a typical helical structure of DNA.
However the effect of the orientation of the dipole moments is
considerably higher with transcendental consequences for the
existence of mobile breathers.
\end{abstract}

%\begin{keyword}
%\sep Discrete breathers \sep Mobile breathers \sep Intrinsic
%localized modes \sep DNA


\begin{multicols}{2}
\narrowtext

\section{Introduction}
%\label{sec:intro}
A great deal of attention has been paid to the interplay between
geometry and nonlinearity in locating problems in recent years.
The relationship between geometry and nonlinearity has an
important role in the functions of some biomolecules, such as DNA,
where the localization of energy has been put forward as a
precursory mechanism of the transcription bubble \cite{Peyrard},
and  moving localized excitations as a method of transporting
information along the double strand \cite{Salerno}.

The fact that  hydrogen bonds that link each pair of bases in DNA
have a finite dipole moment, has brought about the introduction of
models \cite{Christiansen,Mingaleev,Archilla,Cuevas} with long
range dipole--dipole interaction.  Apart from its theoretical
interest, this interaction becomes relevant when the secondary
structure of DNA is considered. The shape of the molecule can
influence the localization and transport properties of energy,
which is thought to play a biological function \cite{DNA}. Some of
these models \cite{Christiansen} study the effects of the
curvature in a chain of nonlinear oscillators using the discrete
nonlinear Schr\"{o}dinger equation. Other models consider
Klein-Gordon systems to study  kinks \cite{Mingaleev}, breathers
in curved chains \cite{Archilla} or breathers with two competing
interactions \cite{Cuevas}. However, all these models with long
range interaction fail to take into account the peculiar
helicoidal structure of the DNA chain, although this has been
considered in some models \cite{Barbi} without the dipole
interaction.

In this paper, we study the effect of  helicity on the properties
of breathers in a Klein--Gordon model with dipole--dipole
interaction. These periodic nonlinear localized oscillations in
discrete systems are very localized excitations that appear as a
consequence of the nonlinearity and discreteness of the system
\cite{Mackay}. They are specially suitable for biomolecules when
considering excitations that involve a few units, that is, far
from the continuous limit. They can be static but, under certain
conditions, also move and transport energy along the system
\cite{Mobile}.

We have found that the introduction of helicity enhances the
stability of static breathers, although this effect is relatively
small for the typical helicoidal structure of the DNA. On the
other hand, the profile of the static breathers and the properties
of moving ones are strongly dependent on the relative orientation
between the dipole moments.

\section{The model}
\label{sec:mod}
 The model is inspired by the primary structure of DNA, with
dipole moments perpendicular to the helix axis, and where the
stretching of the hydrogen bonds within base pairs is described as
a variation of the dipole moments. More detailed justification of
the model can be found in \cite{Cuevas}.

\begin{figure}[h]
\begin{center} \includegraphics[width=6cm]{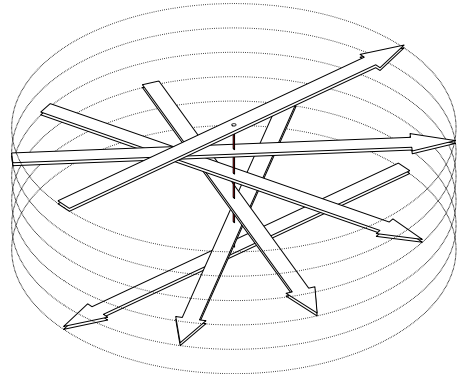}
\end{center}
\caption{Sketch of the model at equilibrium. The arrows represent
the dipoles moments, perpendicular to the helix axis.}
\label{fig1} %helix
\end{figure}

    We denote $\phi_n$ the angle of the n-dipole with respect to a
reference axis perpendicular to the helix axis. Then, the angle
between the nearest neighbouring dipoles is
$\theta_{tw}=\phi_n-\phi_{n-1}$. We have considered this
neighbouring angle constant along the chain, and it will be called
the twisting angle. Thus,  $\phi_{n+m}-\phi_n=m \theta_{tw}$, and,
therefore, $2 \pi/ \theta_{tw}$ dipoles are needed to complete a
turn of screw. In DNA, for example, the twisting angle is
$36^{\mathrm{o}}$ and a turn of screw requires ten base pairs.
Figure \ref{fig1} shows a sketch of the model, where it can be
appreciated that the system of dipoles have an helicoidal
structure.

    In the appropriate dimensionless variables, the Hamiltonian of our
system becomes
\begin{eqnarray}
 H & =& \sum_{n=1}^{N} \bigg(  \frac{1}{2} \dot{u}_n^2+
V(u_n)  \nonumber\\ & & +\frac{1}{2} J
\sum_{m=n-N/2}^{n+N/2}\frac{u_n u_m}{|n-m|^3}\cos [\theta_{tw}
(n-m)] \bigg)\;\; ,
\end{eqnarray}
where $N$ is the number of variables. The variables
$\{u_n\}_{n=1}^N$, where $u_{n\pm N}=u_n$,  represent, in the
context of the Peyrard-Bishop model for DNA \cite{Peyrard}, the
transversal displacements of the two complementary nucleotides in
the n-th pair with respect to the molecular axis. In our model,
they describes the stretching of the dipoles with respect to their
equilibrium length. $V(u_n)$ is the on site potential, which, in
DNA models, describes  the hydrogen bonds linking the two bases,
and the parameter $J$ measures the strength of the long range
dipole-dipole interaction. We have chosen the on--site potential
as the Morse potential, given by
\begin{equation}
V(u_n)=\frac{1}{2}(e^{-u_n}-1)^2
\end{equation}
The reason for this, is that it is a suitable potential for
representing chemical bonds, being asymmetric, with a hard part,
modeling the repulsion between atoms or molecules, and a soft part
that becomes flat, modeling the breakage of the bond.

The dynamical equations become
\begin{equation}
\label{eq:dyn}
    \ddot{u}_n+V'(u_n)+ J \sum_{m=n-N/2}^{n+N/2} \frac{\cos[\theta_{tw} (n-m)]}
{|n-m|^3}u_m=0,
\end{equation}
where n=1\dots N.  To study the linear modes of the system we
replace $V'(u_n)$ in equations~\ref{eq:dyn} with the linear term
$u_n$, which implies that the time has been scaled so that the
linear frequency $\omega_0=1$. Considering solutions of the form
$u_n= e^{iqn-iwt}$ the following dispersion relation is obtained:
\begin{equation}
\label{eq:fonon}
 w_k=\sqrt{1+2J \sum_{m=1}^{N/2}
\frac{\cos(m\,\theta_{tw})}{m^3} \cos(m\,q_k )}
\end{equation}
where $q_k=\frac{2 \pi k}{N}$, with $k=1\dots N$ due to the
periodic boundary conditions.

\begin{figure}
\begin{center}
\includegraphics[angle=-90,scale=0.6]{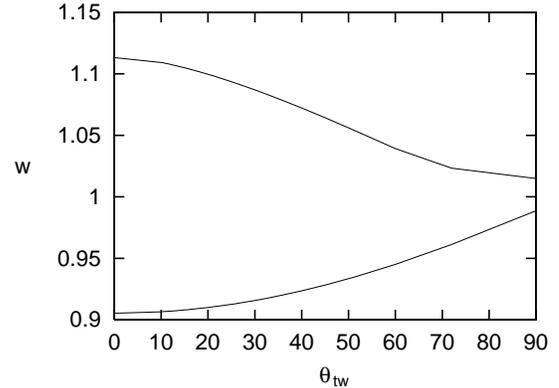}
\end{center}
\caption{Effect of  helicity on the breadth of the linear
spectrum. The curves represent the lower and higher limits of the
phonon band as a function of the twisting angle in degrees for
fixed coupling parameter $J=0.1$. $w$ is in dimensionless units.}
\label{fig2} %linearspectrum
\end{figure}

    The variation of the phonon band with the helicity is
shown in Figure~\ref{fig2}, where the frequencies of the linear
modes are represented as a function of the twisting angle,
$\theta_{tw}$, for a fixed value of the coupling parameter
$J=0.1$. The effect of the twisting is a narrowing of the phonon
band, which will enhance the range of existence and stability of
the breathers. This has been confirmed numerically.

\section{Breather existence and stability}
\label{sec:breathers}
    We have studied the existence and stability of breathers in
this model using the standard numerical methods described in Ref.
\cite{Methods}.

     The Morse potential is a soft potential with
the consequence that the frequency of a breather has to be lower
than the linear frequency $\omega_0=1$. Thus, we have chosen,
$\wb=0.8$ so that the nonlinear effect will be significant but on
the other hand not overtly strong, as the nonlinearities in DNA
are thought to be weak.

\begin{figure}
\begin{center}
\includegraphics[angle=-90,width=8cm]{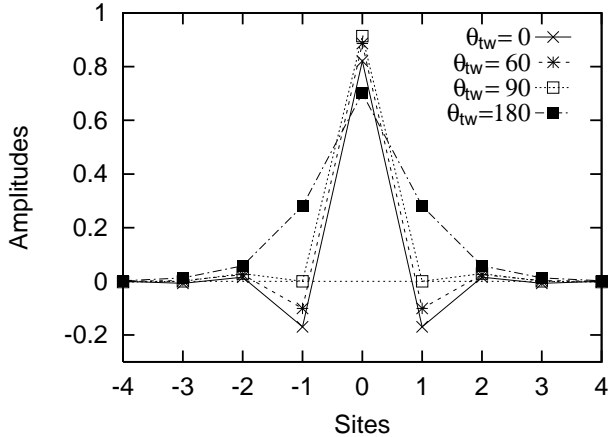}
\end{center}
\caption{Profiles of the one--site breather when twisting is
increased for fixed coupling $J=0.1$ and Morse potential.
Dimensionless units.}
\label{fig3} %:softprofiles
\end{figure}

First of all, the helicity influence the breather profile. As is
shown in figure~\ref{fig3}, for a fixed value of the coupling
parameter $J$, the increase of the twisting angle produces a
transition from a zigzag profile (the nearest neighbour
oscillating in antiphase) to a bell profile (all dipoles
oscillating in phase). This effect follows from the spatial
profile of the phonon state with the lowest frequency since the
breather frequency is below the phonon band, and all the higher
harmonics are way too high to be relevant.  For $\theta_{tw}<
\pi/2$ the interaction is effectively ``antiferromagnetic" which
leads to staggered phonons at the lower band edges. In the same
way for $\theta_{tw}> \pi/2$ a ``ferromagnetic" interaction is
present which leads to a nonstaggered phonon at the lower band
edge.  The breather bifurcates from the lower band edge phonons
and thus retains the property of the phonon structure. For
$\theta_{tw}= \pi/2$ the system separates into two noninteracting
sublattices: even and odd sites. As a result in this case, the
nearest neighbors are at rest and the odd site sublattice remains
unexcited.

\begin{figure}
\begin{center}
\includegraphics[angle=-90,width=8cm]{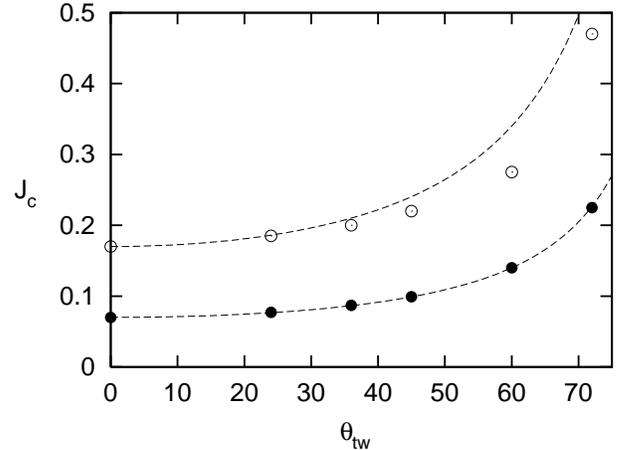}
\end{center}
\caption{Range of stability of the one--site (circles) and two--
sites (full circles) breathers as a function of the twisting angle
in degrees. J$_c$ is the maximum value of the coupling parameter
for which the breather is  stable. The dash lines represent the
values calculate within the NNI approximation. $J_c$ is in
dimensionless units.} \label{fig4}
\end{figure}

    One--site breathers are stable at low coupling as was proved by
Aubry \cite{Aubry}. For any value of the twisting angle
$\theta_{tw}<90^o$ they can be continued from the anticontinuous
limit till $\wb$ enters the phonon band. Just before the breather
disappears, it becomes unstable due to the occurrence of  a
harmonic bifurcation in the evolution of the Floquet eigenvalues.
The increase of the twisting enhances the stability as  is shown
in Fig.~\ref{fig4} (circles). This can be understood if we
consider only the nearest neighbor interaction (NNI). Then the
influence of helicity on the stability of the breathers could be
described by an effective coupling $J_{eff}=J \cos \theta_{tw}$.
The one--site breather without twisting lose its stability for a
coupling value of $J_c^0$. With twisting and only NNI this would
occur for $J_c=J_c^0 / \cos \theta_{tw}$ (dash lines in
Fig.~\ref{fig4}), which concurs with the numerical results.

    The two--site breather, which consist of two
neighboring oscillators excited in phase, is also stable at low
coupling. This can be understood in terms of Aubry's band theory
\cite{Aubry}. When coupling is increased a bubble of instability
appears due to Krein crunches between the phonon band eigenvalues
and a localized eigenvalue of the Floquet operator. If we continue
increasing the coupling  the double breather definitely becomes
unstable due to the occurrence of  a subharmonic bifurcation.
Again, the effect of the twisting is to enlarge the range of
stability toward higher values of the coupling parameter (full
circles in Fig.~\ref{fig4}). This suggests that twisting might be
a way to control the stability of the breathers in real systems.

    We have not considered the two--site breather in antiphase because
it coincides with the one--site breather with zigzag profile,
i.e., the Newton method converges to the same solution if we start
at the anticontinuous limit with one non linear oscillator or with
two nearest neighbor oscillators in antiphase.
    A rather different situation is the one with $\theta_{tw}>90^o$.
First, the one--site breather is always stable until it
disappears. Second, the two--site breather is unstable at low
coupling but  becomes stable just before its extinction. This
behavior has important consequences for the mobility of these
breathers as shown in the next section.

    For the sake of thoroughness we have also studied the effect
of twisting with a hard $\phi^4$ potential $V(u_n)=u_n^2+1/4
u_n^4$, and a breather frequency $\wb =1.2$. Qualitatively the
results are similar except for the fact that breathers with
$\theta_{tw}<90^o$ and breathers with $\theta_{tw}>90^o$ exchange
their properties.

\section{Mobile breathers}
Static  breathers under certain conditions can be moved. The
standard method to move a breather consists in perturbing its
velocities with an spatially antisymmetric vector, called the
marginal mode \cite{Mobile}.  Typically, this method works within
a certain range of parameters near an exchange stability
bifurcation. This occurs when a one--site breather becomes
unstable and a two--site breather does the opposite at a nearby
point.

We have looked for mobile breathers in our system both with a hard
$\phi^4$ potential and with a Morse potential, but we have only
had success with  Morse potential and ``ferromagnetic"
interaction, i.e., $\theta_{tw}>90^o$. In this particular case, we
found a similar situation to a stability exchange and we were able
to move the breather perturbing it with the unstable localized
mode of the two--site breather. This is an interesting result
because this configuration is equivalent to a chain of
antiparallel dipoles twisted $\pi-\theta_{tw}<\pi/2$. In fact, we
can only expect parallel dipoles in synthetic DNA.

A useful concept for describing the breather movement is its
effective mass. If the norm of the perturbation velocity is
$\lambda$, the kinetic energy added to the breather by the
perturbation is $E=\lambda^2/2$. The resulting translational
velocity of the breather, $v$, is found to be proportional to
$\lambda$ \cite{Mobile}. Thus, moving breathers can be considered
as a quasi-particle with a mass of $m^*$, which can be defined
through the relation $m^*v^2/2=\lambda^2/2$.
\begin{figure}
\begin{center}
\includegraphics[angle=-90,width=8cm]{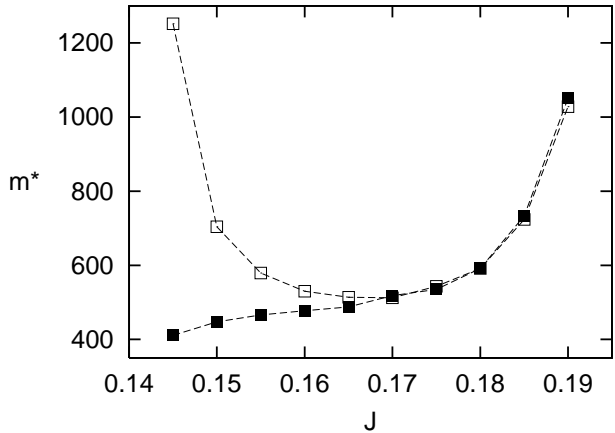}
\end{center}
\caption{Dependence of the effective mass of the mobile breathers
with the coupling parameter for $\theta_{tw}=180^o$. The blank
squares correspond to  mobile breathers obtained from static
one--site breathers. Full squares are obtained from  static
two--site breathers.}
\label{fig5} %:effective mass
\end{figure}
We have studied the dependence of the effective mass, $m^*$, with
the coupling $J$. Figure~\ref{fig5} shows the result   for
antiparallel dipoles ($\theta_{tw}=180^o$). Two different
behaviors were obtained depending on the initial conditions. If we
perturb the two--site breather, we observe that its effective mass
increases monotonically with the coupling (full squares in
Figure~\ref{fig5}). This reflects the fact that the two--site
breather becomes stable with increasing coupling. But if a static
one--site breather is chosen as the initial configuration, a
minimum of $m^*$ appears showing the existence of an optimal value
of the coupling to move this breather (see blank squares in
Figure~\ref{fig5}). We think that this minimum expresses a balance
between the two opposite effects produced by an increase of the
coupling on the stability of the one--site and two--site
breathers. Similar results are obtained for other values
$\theta_{tw}>\pi/2$.

\section{Conclusions}
\label{sec:conclusion}
 We have considered a system of oscillating dipoles with helicoidal
structure in order to study the effect of helicity on the
existence and properties of breathers. This study is motivated by
the helicoidal structure of DNA, and the fact that it can be
described by a reduced dynamic where the only degrees of freedom
are the stretchings of the hydrogen bonds between base pairs,
which have a finite dipole moment. In our model, the helicity
produces a narrowing of the phonon band, and an enlargement of the
range of existence and stability of the breathers, although this
effect is small for a typical helicoidal structure of DNA.

    The effect of the orientation of the dipole moments, i.e., if
the twisting angle is greater or not than 90 degrees, is however
considerably higher. In particular, we have only found mobile
breathers  with a Morse potential and $\theta_{tw}>\pi/2$.
Understanding the necessary conditions to move a breather is still
an opened question today.

\section*{Acknowledgments}
This work has been supported by the European Union under the RTN
project, LOCNET, HPRN--CT--1999--00163. We acknowledge Jes\'us
Cuevas for his useful comments.
\enlargethispage{\baselineskip}

\end{multicols}

\begin{thebibliography}{99}

\bibitem{Peyrard}
M.~Peyrard and A.R.~Bishop, {\em Phys. Rev. Lett.}, 62:2755, 1989.

\bibitem{Salerno}
M.~Salerno and Yu.~Kivshar, {\em Phys Lett A}, 193:263, 1994.

\bibitem{Christiansen}
P.L.~Christiansen, YB.~Gaididei, and S.F.~Mingaleev, {\em J. Phys.
Condens. Matter}, 13(6), 1181 (2001);
%
YuB. Gaididei, S.F.~Mingaleev, and P.L.~Christiansen, {\em Phys
Rev E}, 62:R53, (2000);
%P.L.~Christiansen, Yu~B. Gaididei, M.~Johansson,
%K.{\O}.~Rasmussen, V.K.~Mezentsev, and
%  J.~Juul Rasmussen, {\em Phys. Rev. B}, 57(18):11303--11318, 1998;
S.~F. Mingaleev, P.~L. Christiansen, Yu.~B. Gaididei,
M.~Johansson, and K.~{\O}. Rasmussen, {\em \mbox{J. Biol. Phys.}},
25:41--63 (1999).

\bibitem{Mingaleev}
S.~F. Mingaleev, YB.~Gaididei, E. Majernikova and S. Shpyrko, {\em
Phys Rev E}, 61(4), 4454 (2000).

\bibitem{Archilla}
J.F.R. Archilla, P.L.~Christiansen, S.F.~Mingaleev, and
YuB.~Gaididei, {\em J Phys A: Math. Gen.}, 34:6363, 2001; J.F.R.
Archilla, P.L.~Christiansen, and YuB. Gaididei. {\em Phys Rev E},
65(1):16609, 2002.

\bibitem{Cuevas}
J.~Cuevas, J.F.R. Archilla, YuB. Gaididei, and F.R.~Romero, {\em
Physica D}, 2002, in press.

\bibitem{DNA}
C.~Calladine and H.~Drew.
\newblock {\em Understanding DNA}.
\newblock Academic Press, London, 1992.

\bibitem{Barbi}
M.~Barbi, S.~Cocco, and M.~Peyrard, {\em Phys Lett A}, 253:358,
1999. M.~Barbi, S.~Cocco, M.~Peyrard, and S.~Ruffo, {\em J. Mol.
Biol.}, 24:97, 1999; G.~Gaeta, C.~Reisss, M.~Peyrard, and
T.~Dauxois, {\em Rev. Nuovo Cimento}, 17:1, 1994.


\bibitem{Mackay}
R.S.~MacKay and S.~Aubry, {\em \mbox{Nonlinearity}}, 7:1623, 1994;
S.~Flach and C.R.~Willis, {\em Physics Reports}, 295:181, 1998.

\bibitem{Mobile}
Ding Chen, S.~Aubry, and G.P.~Tsironis, {\em Phys. Rev. Lett.},
77:4776, 1996; S~Aubry and T~Cretegny, {\em Physica D}, 119:34,
1998.

\bibitem{Methods}
J.L.~Marin and S.~Aubry, {\em \mbox{Nonlinearity}}, 9:1501, 1996.

\bibitem{Aubry}
S.~Aubry, {\em \mbox{Physica D}}, 103:201, 1997.

\end{thebibliography}
\end{document}